\documentclass[journal = ancac3]{achemso}
\usepackage{array}
\usepackage{multirow}
\usepackage{amsmath}
\usepackage{amssymb}
\usepackage{graphicx}
\usepackage{color}
\usepackage{booktabs}
\usepackage{dcolumn}
\newcolumntype{d}[1]{D{.}{.}{#1}}
\newcommand\mc[1]{\multicolumn{1}{c}{#1}}
\def\br{{\bf r}}
\def\bp{{\bf r^\prime}}

\def\dd{\rm d}

\title{Nonmonotonic band gap evolution in bent phosphorene nanosheets}
\author{Vojt\v{e}ch Vl\v{c}ek}
\email{vlcek@ucsb.edu}
\affiliation{Department of Chemistry and Biochemistry, University
of California, Santa Barbara California 93106, U.S.A.}

\author{Eran Rabani}
\email{eran.rabani@berkeley.edu}
\affiliation{Department of Chemistry, University of California and Materials Science
Division, Lawrence Berkeley National Laboratory, Berkeley, California
94720, USA}
\affiliation{The Raymond and Beverly Sackler Center for Computational Molecular
and Materials Science, Tel Aviv University, Tel Aviv, Israel 69978}

\author{Roi Baer}
\email{roi.baer@huji.ac.il}
\affiliation{Fritz Haber Center for Molecular Dynamics, Institute of Chemistry,
The Hebrew University of Jerusalem, Jerusalem 91904, Israel}

\author{Daniel Neuhauser}
\email{dxn@ucla.edu}
\affiliation{Department of Chemistry and Biochemistry, University of California,
Los Angeles California 90095, U.S.A.}

\begin{document}
\begin{abstract}
Nonmonotonic bending-induced changes of fundamental band gaps and quasiparticle energies are observed for realistic nanoscale
phosphorene nanosheets. Calculations using stochastic many-body perturbation theory (s$GW$) show that even slight curvature causes significant changes in the electronic properties. For small bending radii ($<4$~nm) the band-gap changes from direct to indirect. The response of phosphorene to deformation is strongly anisotropic (different for zig-zag vs. armchair bending) due to an interplay of exchange and correlation effects. Overall, our results show that fundamental band gaps of phosphorene sheets can be manipulated by as much as $0.7$~eV depending on the bending direction.
\end{abstract}

\maketitle
\section{Introduction}
Since its discovery less than a decade
ago~\cite{liu2014phosphorene,koenig2014electric,li2014black}, single
layer phosphorene attracted much attention due to its unique
electronic and mechanical properties. Its fundamental band gap ($E_g$)
can be tuned by increasing the number of stacked
monolayers~\cite{tran2014layer,qiu2017environmental,li2017direct} or
by chemical doping~\cite{kim2015observation}, and spans a wide range
of values, from $E_g=1.88$~eV in single layer phosphorene to
$E_g=0.3$~eV in the bulk. The unique mechanical
properties~\cite{wei2014superior} along with high room temperature
mobilities (around $1,000 \mbox{cm}^2/\mbox{s V}$)~\cite{li2014black}
make phosphorene a promising candidate for fabrication of next generation
flexible
nanoelectronics~\cite{Hone2014,li2014black,xia2014rediscovering,buscema2014fast,zhu2016black}
nanophotonics~\cite{xia2014two}, and ultrasensitive
sensors~\cite{koppens2014photodetectors, kou2014phosphorene}.

Understanding the interplay between the electronic and mechanical
properties is central for future technological developments.  Indeed,
significant progress has been made in describing the role of
strain. Density functional theory (DFT) calculations predict a
decrease in the band gap as a result of the application of uniaxial
strain, which ultimately results in a direct-to-indirect band gap
transition~\cite{rodin2014strain,peng2014strain}. However, DFT is not
a good proxy for quasiparticle energies~
\cite{gross1990density,martin2016interacting}.  The case of bent
phosphorene is even more challenging, since investigation of bending
effects naturally precludes the use of periodic boundary
conditions. Thus, studies so far have been limited to narrow
(quasi-1D) phosphorene nanoribbons~\cite{yu2016bending} within DFT,
indicating charge localization and formation of in-gap states for
extreme bending conditions (radii $R<1.3$~nm). These bending
scenarios are very challenging experimentally.

Ab-initio many-body perturbation theory in the $GW$
approximation~\cite{hedin1965new,aryasetiawan1998gw,hedin1999correlation} yields accurate predictions for
quasiparticle energies.  Its cost was prohibitive, however, so $GW$ was only feasible for small and medium sized systems \cite{deslippe2012berkeleygw,govoni2015large}.  Luckily, the costs are drastically reduced by a new stochastic approach to simulating GW, labeled StochasticGW or just s$GW$~\cite{neuhauser2014breaking,vlcek2017stochastic,vlcek2018swift}, which is a part of a general stochastic paradigm~
\cite{neuhauser2012expeditious,Baer2013,Neuhauser2014,Arnon2017,Chen2018,Fabian2018}.
 s$GW$ is sufficiently efficient that it is less expensive than the underlying DFT stage, and this makes it possible to treat systems with thousands of electrons or more~\cite{vlcek2017stochastic,vlcek2018swift}.  We employ
here s$GW$  for calculating quasiparticle (QP)
energies for a series of large ($2.9\times4.3$~nm) phosphorene
nanosheets (PNS). 

The PNS are subject to bending with radii between
$1\mu$m and $2$~nm -- a range that can be realized
experimentally~\cite{levy2010strain,tang2014nanomechanical}. Thus, it
is possible to directly map the evolution of band gaps with deformation of
a 2D material. We discover here that even a small sample curvature affects
the QP energies and that DFT severely underestimates the response to
bending. Further, irrespective of the direction of bending, we find an
interesting crossing of the lowest unoccupied states leading to a
change of character of the gap for radii $<4$~nm. The PNS response is
strongly anisotropic and is governed by nontrivial interplay of
exchange and correlation effects.  Our results predict that under
realistic conditions, the QP gap can be manipulated solely by deformation
by as much as $0.7$~eV.

\section{Theory and Methods}
Fundamental band gaps are defined as differences between ionization
potential and electron affinity, which correspond to quasiparticle
(QP) energies of the highest occupied ($\varepsilon^{QP}_H$) and
lowest unoccupied states ($\varepsilon^{QP}_L$), i.e.:
\begin{equation}
E_g =  \varepsilon^{QP}_L -\varepsilon^{QP}_H.
\end{equation}
While density functional theory yields a set of eigenstates and
corresponding eigenvalues, those cannot be interpreted as QP
energies~\cite{gross1990density}. Indeed, DFT eigenvalue differences
severely underestimate true band gaps~\cite{martin2016interacting}. A
solution is to calculate $\varepsilon^{QP}$ through many-body
perturbation theory with Kohn-Sham (KS) DFT as a starting
point~\cite{hybertsen1986electron,aryasetiawan1998gw,hedin1999correlation}.

KS eigenvalues ($\varepsilon^{KS}$) contain contributions from kinetic
energy and Hartree, ionic and a mean field exchange-correlation (xc)
potential energies. The QP energy is obtained by replacing the xc term
($v_{xc}$) by exchange ($\Sigma_X$) and polarization self-energies
($\Sigma_P$):
\begin{equation}\label{eqp_eq}
\varepsilon^{QP} = \varepsilon^{KS} - {v}_{xc} + \Sigma_X + \Sigma_P\left(\varepsilon^{QP}\right).
\end{equation}
The exchange contribution is 
\begin{equation}\label{eq:X}
\Sigma_X = -\sum_j^{N_{occ}} \iint \phi\left(\br\right) \phi_j\left(\br\right) \frac{1}{\left| \br - \bp\right|} \phi_j\left(\bp\right)  \phi\left(\bp\right) \dd \br \dd \bp,
\end{equation}
where $\phi$ is the orbital for which $\varepsilon^{QP} $ is evaluated
and the sum extends over all $N_{occ}$ occupied states. $\Sigma_P$ is a
dynamical quantity describing the polarization of the density due the QP. Note that Eq.~(\ref{eqp_eq}) is a fixed point equation,
where $\Sigma_P$ is evaluated at the frequency corresponding to
$\varepsilon^{QP}$.

The self-energy terms are computed using s$GW$, which, as mentioned, scales nearly linearly with number of electrons
and allows to compute $\Sigma$ for extremely large systems with
thousands of atoms~\cite{vlcek2018swift}. While the $GW$ approximation
should in theory by solved by a self-consistent set of Hedin's
equations~\cite{hedin1965new}, it is common practice to use a one-shot
correction ($G_0W_0$), in which the self-energy is based on underlying
KS Hamiltonian. This is however insufficient in many
cases~\cite{martin2016interacting}. We thus rely on a partially
self-consistent $\bar{\Delta}GW$ approach~\cite{vlcek2018simple}
which is a simple post-processing step on top of $G_0W_0$ and yields
band gaps in excellent agreement with
experiment~\cite{vlcek2018simple}.

\section{Results}
We investigated the effects of bending on a set of phosphorene
nanosheets derived from the experimental structure of bulk black
phosphorus~\cite{cartz1979effect}. PNS were constructed from a
$10\times10$ single sheet supercell passivated with hydrogen atoms.
We relaxed the interatomic positions using a reactive force field
developed for low dimensional phosphorene
systems~\cite{xiao2017development}. First principles geometry minimization, e.g., with DFT, is too expensive due to the size of the system.
The relaxation was performed such that the phosphorus
atoms that are on the straight edge were fixed and the structure
optimized within the LAMMPS
code~\cite{plimpton1995fast,aktulga2012parallel}.  

A ground state DFT
calculation was performed using a real-space grid representation,
ensuring (through the Martyna-Tuckerman
approach~\cite{martyna1999reciprocal}) that the potentials are not
periodic. The exchange-correlation interaction was described by the local
density approximation (LDA)~\cite{PerdewWang} with Troullier-Martins
pseudopotentials~\cite{TroullierMartins1991}.  With a kinetic energy cutoff
of $26 E_h$ and $0.6 a_0$ real-space grids-spacing the Kohn-Sham eigenvalues were converged to $<$10~meV.  

Many-body
calculations were performed using the StochasticGW code~\footnote{Code is
  available at http://www.stochasticgw.com/} with $40,000$ fragmented
stochastic bases. Only quasiparticle energies were computed, while we
kept the DFT orbitals unchanged. The dynamical part of the self-energy
was computed using $8$ stochastic orbitals in each stochastic sampling
of $\Sigma_P$ using the random-phase approximation (i.e.,
time-dependent Hartree) and with a propagation time of $100$ atomic
units. The total number of stochastic samples was varied to reach a statistical error of $\le$ 0.02 eV for the QP energies (typically 1,200
samples).

\subsection{Planar phosphorene nanosheet}
\begin{figure}
\includegraphics[width=0.5\textwidth]{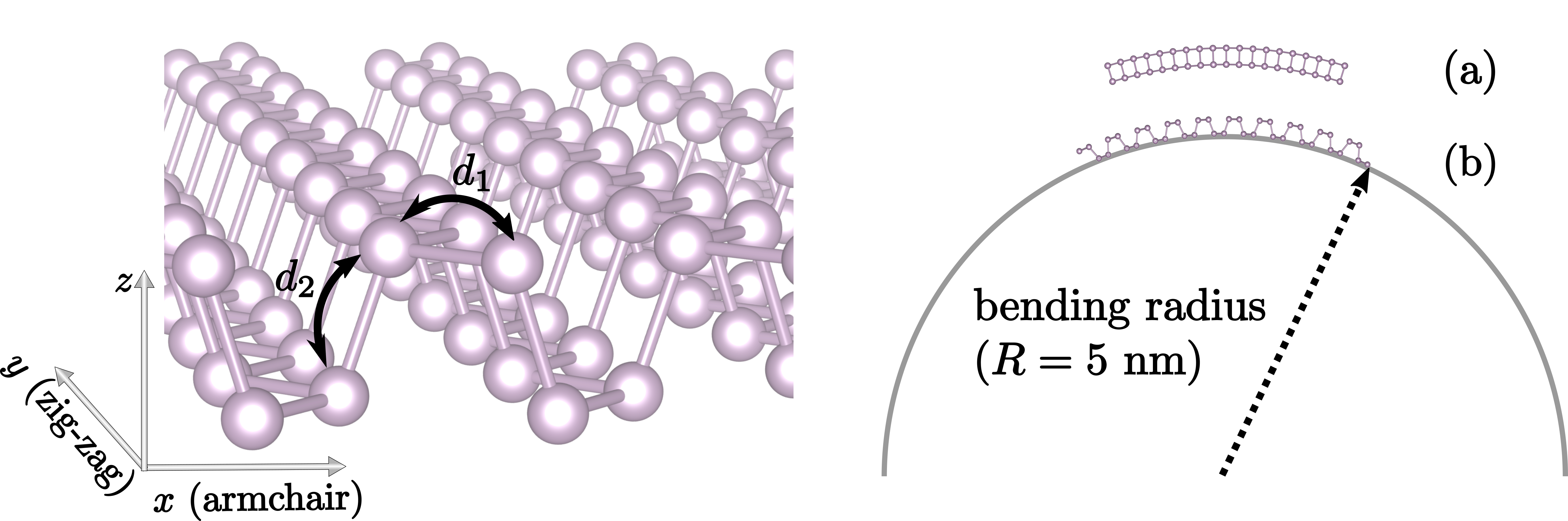}
\caption{Phosphorene is characterized by puckered honeycomb
  lattice with ridges along the armchair ($x$)
  axis. Two nearest neighbor distances are denoted in the left panel
  ($d_1$ and $d_2$).  Bending of PNS is illustrated on the right for
  bending radius $R=5$~nm; PNS bent along zig-zag and armchair axes
  are on the top denoted as (a) and (b), respectively. }\label{fig:struct}
\end{figure}

Ideal phosphorene geometry has a puckered honeycomb structure with two
distinct in-plane directions: armchair ($x$) and zig-zag ($y$) as
shown in Fig.~\ref{fig:struct}. The characteristic ridges in the structure are along the zig-zag direction. Each phophorus atom has two
nearest neighbor distances $d_1$ and $d_2$ constituting a ridge. In
two extreme scenarios, the bending axis is either along the $x$ (armchair) or $y$ (zig-zag)
directions (Fig.~\ref{fig:struct}), resulting in a nonuniform inter-atomic distances.

To focus our investigation purely on the effect of PNS bending, we
first construct an ideal phosphorene monolayer with dimensions
$4.3\times2.9$~nm along the armchair and zig-zag directions with $1,958$
valence electrons.  The arrangement of the P atoms is identical to a
layer of the periodic P crystal~\cite{cartz1979effect} and thus our results
can be compared to previous calculations for an infinite 2D
systems.

We find that for a planar PNS, one-shot $G_0W_0$ predicts
a quasiparticle band gap of $E_g=2.23\pm0.04$~eV. This is larger by
$\sim0.2$~eV than $E_g$ for bulk systems~
\cite{liang2014electronic,tran2014layer,qiu2017environmental}. Self-consistency
($\bar{\Delta}GW$) further increases the fundamental band gap to
$E_g=2.47\pm0.04$~eV.  Our $\bar{\Delta}GW$ result overlaps a previous study of infinite 2D sheets of phosphorene at the $G_1W_1$ level (obtained in first iteration to
self-consistency)~\cite{tran2014layer} but is larger by $0.17$~eV
than a similar self-consistent treatment  ($GW_0$) for
bulk~\cite{rasmussen2016efficient}. The larger fundamental gap indicates that the large PNS considered here is still slightly influenced by quantum
confinement, but to a much smaller degree than several small systems that were previously studied by DFT~\cite{allec2016inconsistencies,yu2016bending}.  This shows the strength of s$GW$, which provides reliable results for quasiparticle energies of extended systems.

\begin{figure}
\includegraphics[width=0.45\textwidth]{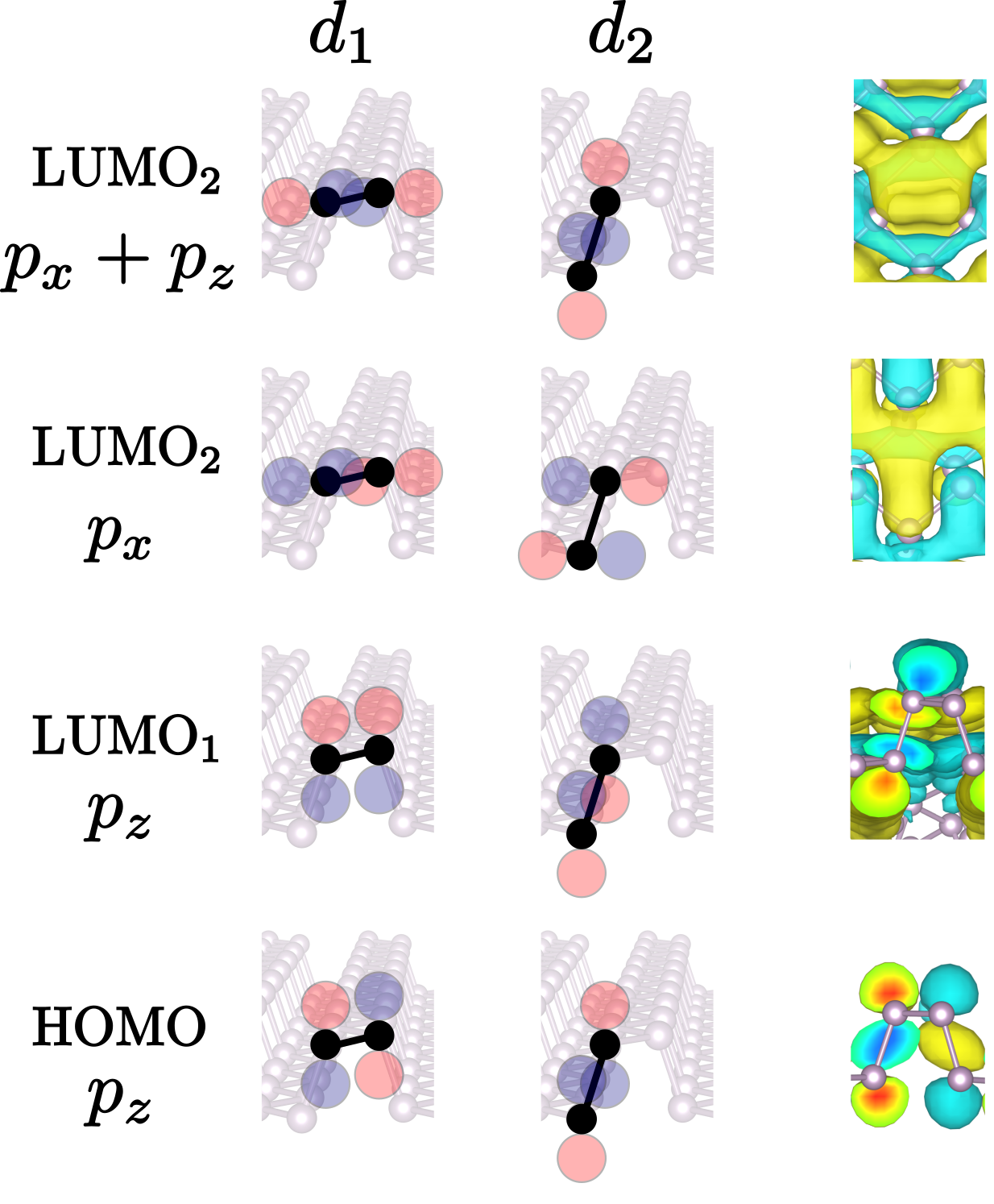}
\caption{Orbital character for the band-edge states. Simplified
  overlaps of two nearest neighbors with interatomic distances $d_1$
  and $d_2$ are depicted separately for
  clarity. Identical colors in the $p$-orbital lobes correspond to a bonding overlap, distinct colors depict an anti-bonding overlap. The rightmost column shows details of the orbital
  isosurfaces. The character of the HOMO state does not change with
  bending. LUMO$_1$ is the lowest unoccupied state for radii
  $R\ge4$~nm. LUMO$_2$ (see text for reference) has either an anti-bonding $p_x$ character (for
  $R<4$~nm in zig-zag bending) or mixed $p_x + p_z$ character
  (for $R<4$~nm in armchair bending).  }\label{fig:orbs}
\end{figure}

By inspecting the nature of individual states (Fig.~\ref{fig:orbs}),
we find that the valence band maxima and the conduction band minima have a $p_z$ orbital character. In a simplified picture, the $p$ orbitals are
centered on each P atom and their hybridization forms bonding and
anti-bonding states. This is qualitatively shown in the HOMO and LUMO in Fig.~\ref{fig:orbs}. Since bending (discussed later)
changes the orbital ordering, we denote the lowest unoccupied
state in a planar system as LUMO$_1$, for clarity.

\begin{figure}
\includegraphics[width=0.35\textwidth]{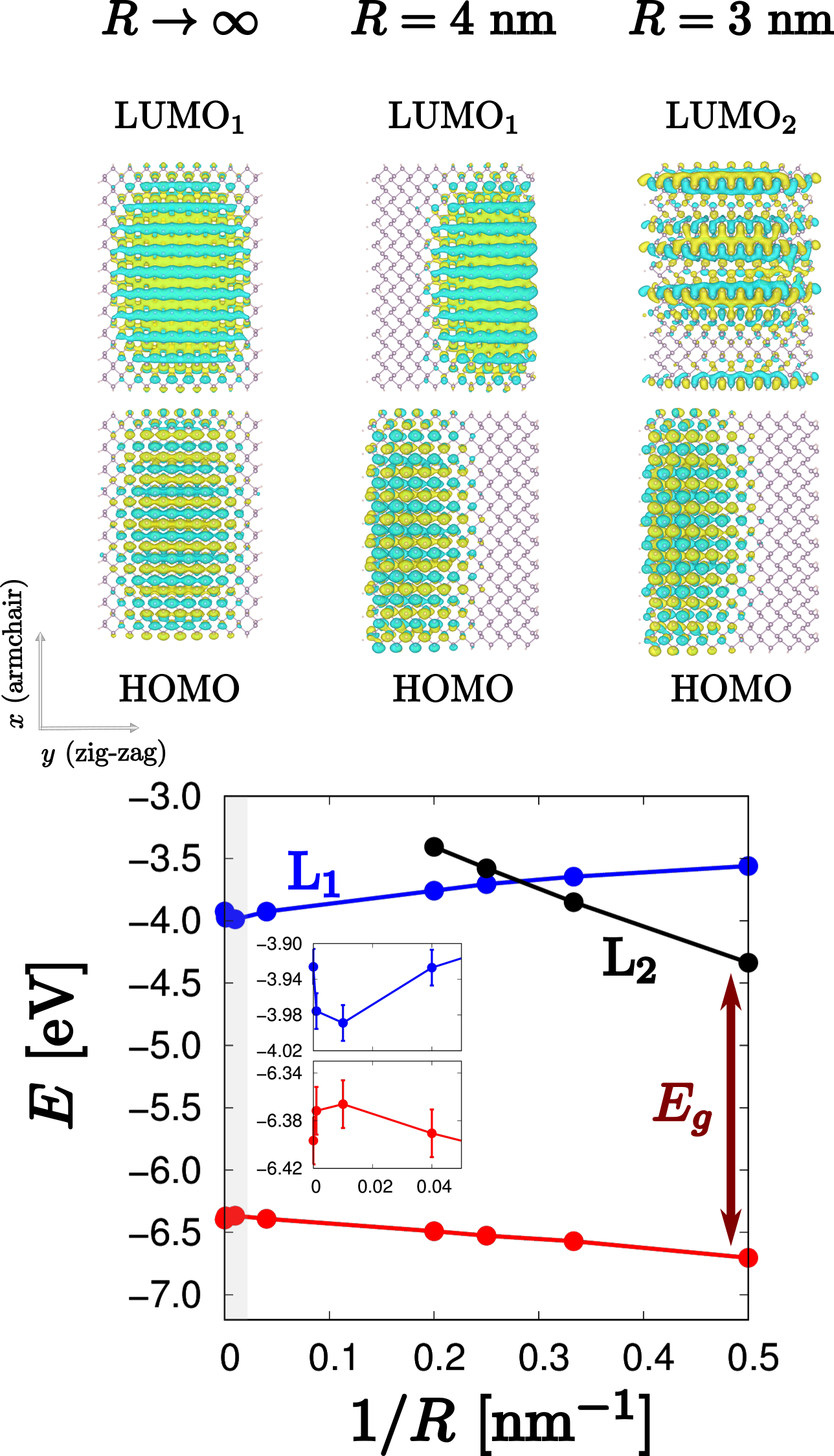}
\caption{Top: Orbital isosurfaces for bending along the zig-zag axis; the phase of the wave-function is distinguished by its color. Bottom: QP energies in zig-zag bending. The stochastic
  error on each point is smaller than the symbol size, and is explicitly shown in the inset. The line is a
  guide for the eye. The HOMO state is shown in red, LUMO$_1$ (denoted
  L$_1$) in blue and LUMO$_2$ (denoted L$_2$) in black. The
  fundamental band gap $E_g$ is shown for $R=2$~nm,  and the specific values are reproduced in Table~\ref{tab:gaps}. Note that LUMO$_2$ is identified for
  $R=5$ as the fifth state above LUMO$_1$ (for clarity we do not depict
  intermediate states).  }\label{fig:orbs_PNS_zz}
\end{figure}

\begin{figure}
\includegraphics[width=0.35\textwidth]{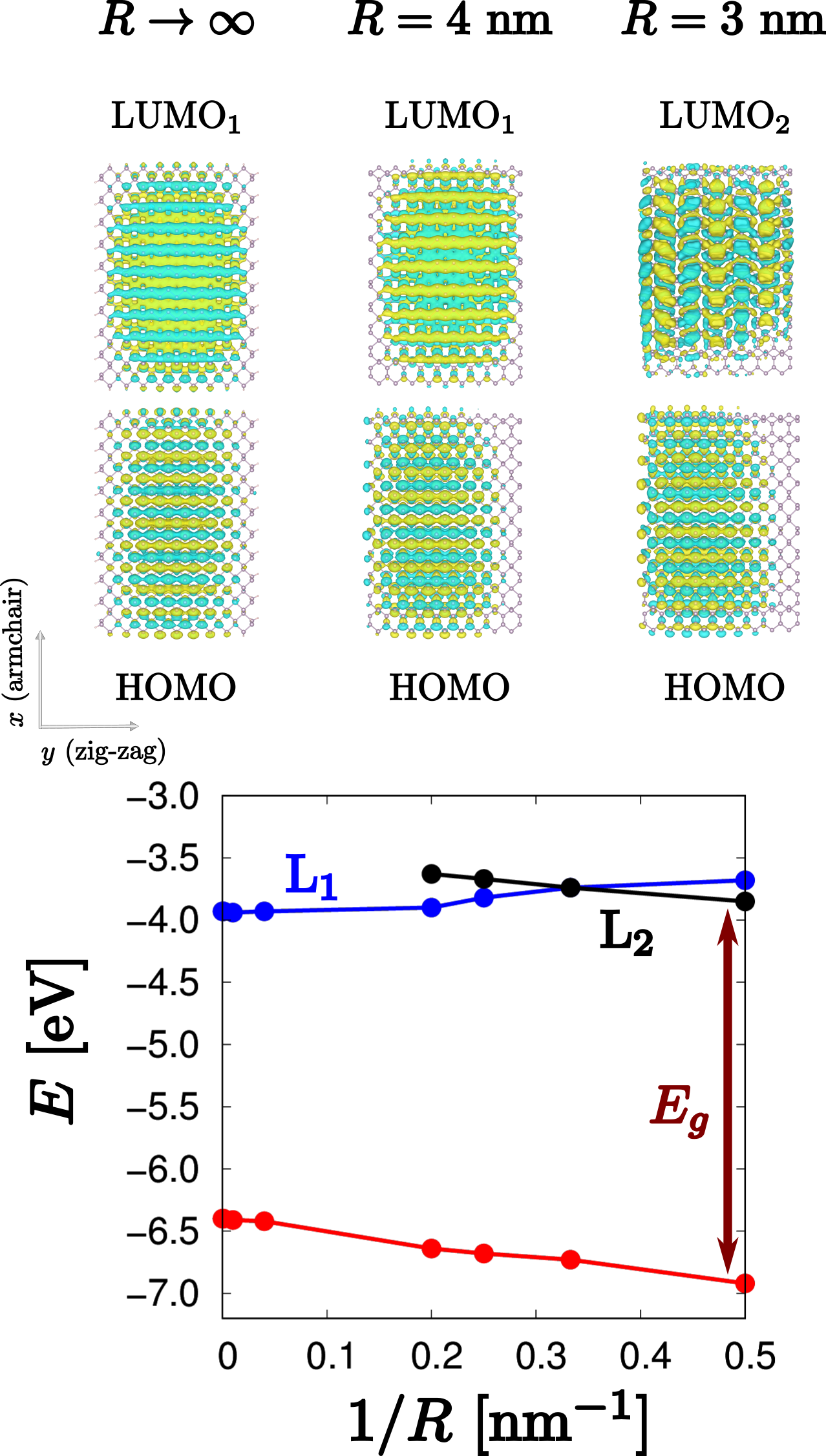}
\caption{Top: Orbital isosurfaces for bending along the armchair axis. Bottom: QP energies for PNS
  bent along the armchair direction. }\label{fig:orbs_PNS_ac}
\end{figure}

Both HOMO and
LUMO$_1$ states are strongly delocalized around the center of the PNS and
extend to the edges along the armchair ($x$) direction
(see the left isosurfaces in Figs.~\ref{fig:orbs_PNS_zz} and \ref{fig:orbs_PNS_ac}, in the limit $R\rightarrow \infty$). The delocalization along the armchair direction is associated with an effective mass that is $7$-times lower along the armchair direction compared to the zig-zag direction~\cite{peng2014strain}. The orientation and phase of
the orbitals does not change markedly when translating by a unit-cell
vector along $x$ or $y$ direction. This indicates that both HOMO and
LUMO$_1$ are in-phase, consistent with previous calculations for
bulk~\cite{rodin2014strain,peng2014strain,allec2016inconsistencies},
supporting a direct band-gap material.

We also performed calculations for phosphorene nanosheets relaxed with a
reactive force-field which was tuned to reproduce the elastic
properties of phosphorene~\cite{xiao2017development}. Relaxation
affects mainly atoms at the edge and shortens the $d_2$ distance by
0.03\AA. As a result, the $d_1$ and $d_2$ bond lengths are almost
identical, leading to stabilization of the $p_x$ character at the expense
of $p_z$ states~
\cite{rodin2014strain,peng2014strain,li2017direct,qiu2017environmental}, signifying that the particular ordering of electronic states in phosphorene is very sensitive to the geometry.  In the next subsection, we illustrate however that while the quasparticle band gaps change dramatically and qualitatively by bending, this does not depend on the precise
geometry of the monolayer.

\begin{figure}
\includegraphics[width=0.35\textwidth]{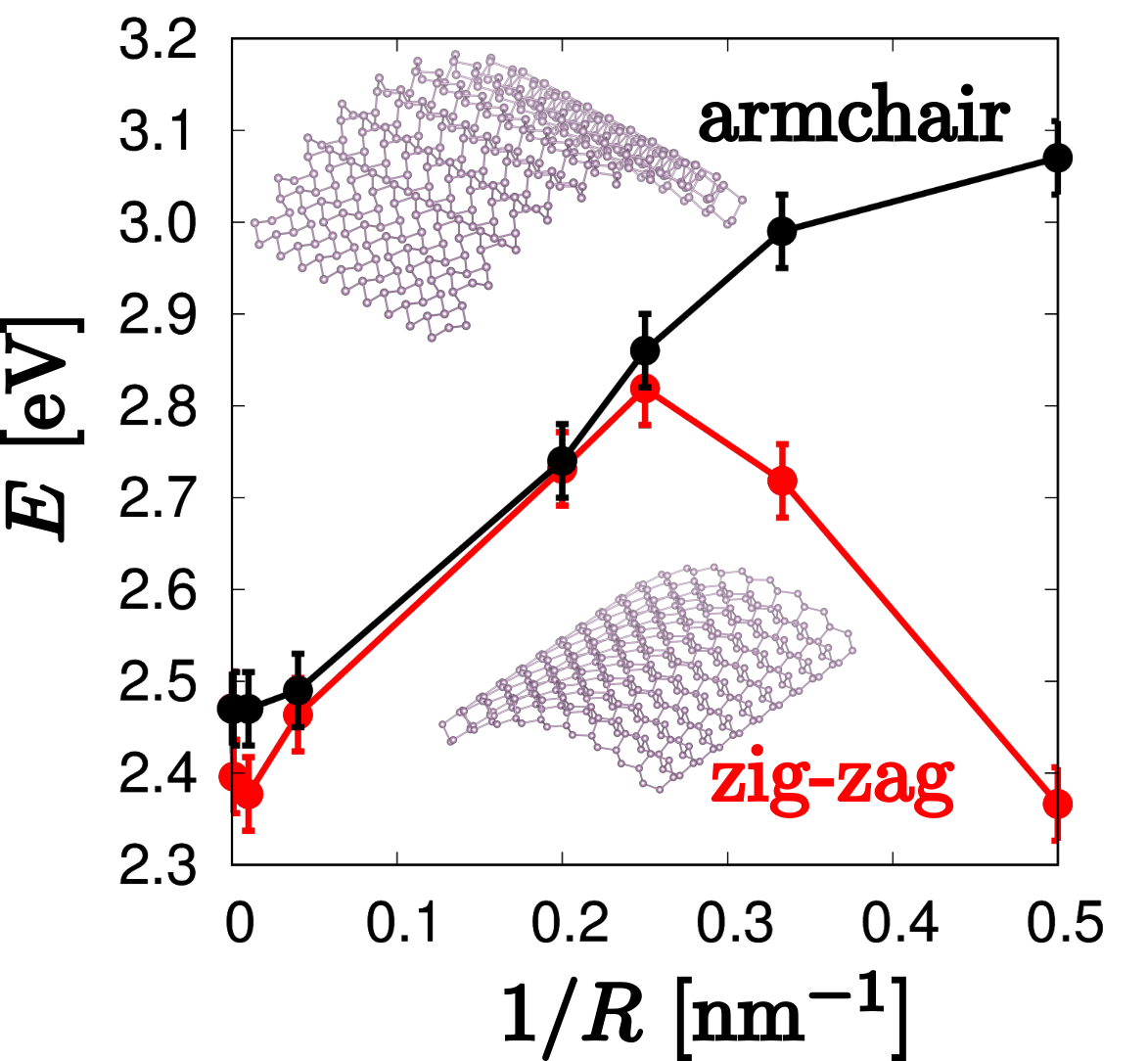}
\caption{Fundamental band gaps ($E_g$) of phosphorene sheets bent along the armchair (black) and zig-zag (red) directions. Bent structures for $R=2$~nm are illustrated in the insets: top for armchair bending, bottom for zig-zag. Error bars show the stochastic errors. The lines are guides for the eye. }\label{fundgaps}
\end{figure}

\subsection{Bent phosphorene nanosheet}
\subsubsection{Zig-zag bending}
Even the slightest deformation along the zig-zag direction (the $y$-axis of the sheet is bent, see Fig.~\ref{fig:struct}) results in changes in
quasiparticle energies (Fig.~\ref{fig:orbs_PNS_zz}). For large bending
radii between $1$~$\mu$m and $100$~nm (see inset in the bottom panel of Fig.~\ref{fig:orbs_PNS_zz}) the HOMO energy increases and
the LUMO$_1$ energy decreases with bending radius. The fundamental band
gap consequently drops by $0.10\pm0.04$~eV; this is clearly seen in Fig.~\ref{fundgaps} which shows the evolution of $E_g$ with $1/R$. Note that for such large
$R$ the change in the atomic positions is rather small, $<0.2\%$.  This effect is seen only for nanosheets bent along the zig-zag direction, but irrespective of the direction of the bending along the $z$-axis (whether they are bent up or down). It is likely related to left-right symmetry breaking (along the $y$-axis) that allows the HOMO and LUMO$_1$ orbitals to shift towards the edges, as discussed in the next section.

If the bending radius is further decreased ($100$~nm$> R > 4$~nm), the
HOMO and LUMO$_1$ states gradually shift even more towards opposite edges
parallel to the bending axis but remain extended along the armchair
direction (Fig.~\ref{fig:orbs_PNS_zz}). The energies of both states depend nearly
linearly on the inverse bending radius as shown in bottom panel of
Fig.~\ref{fig:orbs_PNS_zz}. The HOMO decreases with a slope of
$-0.71$~eV~nm, but LUMO$_1$ increases with slope of $1.05$~eV~nm. As
a result, the fundamental band gap opens up with decreasing bending radius. 
For $R=4$~nm the band gap rises to  $2.82\pm0.02$~eV, significantly larger than the band gap for planar
PNS ($2.47\pm0.04$~eV), as clearly shown in Fig.~\ref{fundgaps}. 

For very small bending radii ($<4$~nm), we observe a transtion in the order of $\mbox{LUMO}_1$ and $\mbox{LUMO}_2$. The latter is nearly triply degenerate and becomes the lowest unoccupied orbital. As a results, the band gap decreases with bending radius and for $R=2$~nm  the  band gap is $E_g= 2.37\pm0.04$~eV, i.e., even lower than the bulk value (cf.~Fig.~\ref{fundgaps}). 

At any bending radius, both HOMO and LUMO$_1$ retain their $p_z$ character.  Similarly, LUMO$_2$, which is triply degenerate, has a $p_x$ character (and, as mentioned, dips below LUMO1 when the bending radius is smaller than 4nm). Note that LUMO$_2$ is characteristically delocalized over the ridges (i.e., along the zig-zag direction).  Some examples of orbital isosurfaces (including one of the three LUMO$_2$ states) are shown in Fig.~\ref{fig:orbs_PNS_zz}. Specifically, for the the outer (dilated) surface of the phosphorene nanosheet, the neighboring P atoms exhibit anti-bonding $p_x$ overlap. In contrast, for atoms on the inner (contracted) surface the overlap has a bonding character. We further note that the LUMO$_2$ orbital is localized on every other ridge along the armchair direction, i.e., it has periodicity twice as long. This indicates that the fundamental band gap becomes indirect. The preceding discussion and the plot in Fig.~\ref{fig:orbs_PNS_zz} were for one of the LUMO$_2$ states, but the two other LUMO$_2$ states behave similarly.

\subsubsection{Armchair bending}

The PNS is also sensitive to bending along the armchair direction, as summarized in Fig.~\ref{fig:orbs_PNS_ac}, but the overall trends are quite different. Now, both HOMO and LUMO$_1$  shift negligibly towards the armchair edges. Unlike zig-zag bending, $E_g$ remains practically constant till $R$ is lower than 100~nm and increases when the system is further bent. This is quantitatively shown in Fig.~\ref{fundgaps}.  

The increase in the fundamental gap for $R<100$~nm is mainly due to a shift of the HOMO that decreases linearly with slope of $-1.05$~eV~nm. This slope is 50\% larger in magnitude than in deformation along the zig-zag axis.  For radii $<4$~nm, we also observe a crossing of the two unoccupied states, as was the case for zig-zag bending. The shape of LUMO$_2$ in this armchair bending case is, however, quite different.  LUMO$_2$ has now a mixed   $p_z$ and $p_x$ character and its phase is roughly four times larger than a single unit-cell, while HOMO and LUMO$_1$ have the same spatial periodicity as the ionic structure. This suggests that for highly bent systems, the band gap is indirect.   We further observe that the QP energy of  LUMO$_2$  decreases slowly ($-0.66$~eV~nm). Consequently, the band gap opens with a mild positive slope ($0.39$~eV~nm). For $R=2$~nm, we obtain $E_g = 3.08\pm0.02$~eV, which is $0.7$~eV larger  than for a PNS bent by a similar amount along the zig-zag direction, and 0.6~eV larger than for a planar phosphorene. The QP band gaps are shown in Fig.~\ref{fundgaps} and, for selected radii, in Table~\ref{tab:gaps}.

\subsubsection{Force-field-optimized bent structures}

We have also computed band gaps for relaxed phosphorene nanosheets with $R=4$ and $2$~nm. The geometries were relaxed keeping the outermost edge P atomic positions fixed. 
As we mentioned in the previous section, with force-fields relaxation even a planar ($R\to \infty$) structure the lowest LUMO has a $p_x$ character.  With force-field relaxation, bending along the zig-zag direction does not lead to state crossing. The LUMO keeps a $p_x$ character, and its energy decreases with bending radius.

In contrast, when a force-field relaxed structure is bent along the armchair direction, the $p_z$-type orbital becomes a tiny bit more stable than the $p_x$ one.  The difference is so small that both LUMO states are practically degenerate. 

In spite of the difference in state character between the idealized and force-field optimized structures, they both show the same difference (0.7eV) between the band-gaps of zig-zag and arm-chair bent structure at $R$=2nm. Therefore, the precise state ordering depends on geometrical details, but the overall response to bending is highly anisotropic.

\section{Discussion}

\subsubsection{Small curvatures}

We now turn to analyze the results, and start with large $R$.  Here, the behavior described in the previous section is remarkable.  Recall that upon a tiny change of curvature in the zig-zag direction (from $R\to \infty$ to $R\ge100$~nm), the band-gap decreases by about $0.1$~eV (Fig.~\ref{fundgaps}).  This is not a big change compared with the changes at $R\sim2-4$~nm, but it occurs with only a tiny modification of geometry.  Further, this effect was not seen in DFT calculations.

To understand this zig-zag induced $0.1$~eV change, we need to first recall that the system is highly anisotropic. Fig.~\ref{fig:orbs_PNS_zz} shows that HOMO and LUMO$_1$ are strongly confined only along the armchair direction.  This is consistent with the highly anisotropic effective masses of electrons and holes ($0.16/0.15 m_e$ and $1.24/4.92 m_e$ along the armchair and zig-zag directions, respectively for electrons/holes~\cite{peng2014strain}). Upon even a tiny bending (i.e., at any finite $R$), the HOMO and LUMO$_1$ can easily migrate to the sides, as shown in Fig.~\ref{fig:orbs_PNS_zz}.  The energy required to localize the orbitals along the $y$-axis is negligible due to the large effective mass along the zig-zag direction.

In a previous DFT study~\cite{peng2014strain}, a large amount of strain (4\%) was required to induce the same size of band-gap modification ($0.1$~eV). This is much larger than the strain in small-curvature bending (for $R=100$~nm the strain is only 0.02\% along the $y$ direction). Further, the $0.1$~eV induced zig-zag bending effect is only observed in $GW$.  The underlying DFT calculations do not show eigenvalue modifications for such tiny bending (i.e., $R>100$~nm). This mechanism suggests that even small curvature of real finite samples may change significantly the fundamental gaps.

\subsubsection{Large curvatures}

We now turn to large-curvature bending, with $R$ between $100$~nm and $2$~nm.  In DFT the QP energies change is small ($0.1$~eV or less).  In $GW$, however, the changes are significant, as we mentioned in the previous section, and as also shown quantitatively in Table~\ref{tab:gaps}.

 The change of QP energies in $GW$ comes from two sources: exchange ($\Sigma_X$) and polarization ($\Sigma_P$).  Exchange is overall stronger, but we find many cases where the polarization is almost as big in magnitude. To analyze the relative contributions, we fit the exchange-only contribution by a tight-binding-like expression:
\begin{equation}\label{SX_approx}
\Delta\left(\Sigma_X\right) \simeq 
   O_{1}  \Delta \left(\frac{1}{d_1}\right)  
+ O_{2}  \Delta \left(\frac{1}{d_2}\right)  .
\end{equation}
 Here,  $O_1$ and $O_2$  are fitted parameters, while $d_1$ and $d_2$ are the \emph{average} interatomic distances and $\Delta$ refers to the change relative to the planar structure.

Due to the finite thickness of a single PNS, atoms on the ``outside'' and ``inside'' experience slightly different curvature and hence the interatomic distances vary. This is reflected in Eq.~\ref{SX_approx} by considering an average interatomic distance.
Upon bending, the average distances increase as the dilatation of the outer-surface distances is larger than the compression of the inner surface ones, so $1/d$ decreases. For armchair bending both $d_1$ and $d_2$ change (the former about 10-times as much as the latter); for the zig- zag bending only $d_1$ changes. \footnote{For the maximum bending, i.e., $R=2$~nm, we achieve the largest change of the interatomic distance: on average $d_1$ is elongated by 9\% for bending along both $x$ and $y$ axes, while $d_2$ changes merely by  1\% and happens only for bending along the zig-zag direction. } 

In our model (Eq.~\ref{SX_approx}), the bonding orbitals stabilize $\Sigma_X$: they  have a negative value of $O_{1,2}$  and upon shortening of interatomic distances (i.e., when $\Delta\left(1/d_{1,2}\right) >0$) the exchange self-energy becomes more negative (i.e., $\Delta\left(\Sigma_X\right) < 0$) . In contrast, the anti-bonding orbitals destabilize the QP energy as the atoms become closer, i.e., they are associated with positive values of $O_{1,2}$.

Table~\ref{tab:sigma_x_model} contains the fitted $O_{1,2}$ coefficients for the HOMO, LUMO$_1$ and LUMO$_2$ (the latter for zig-zag bending during which LUMO$_2$ has a $p_x$ character).  Note the reverse signs of $O_1$ and $O_2$ for HOMO and LUMO$_1$ (first two rows of Table~\ref{tab:sigma_x_model}). The opposite signs indicate distinct bonding/anti-bonding characters along $d_1$ and $d_2$ for the two band-edge states. As mentioned in the previous paragraph, bending causes (on average) $d_1$ to increase much more than $d_2$, i.e., the $O_2$ contribution results in smaller quantitative changes.

During bending along both directions, HOMO becomes less destabilized by the ``anti-bonding interaction'' along $d_1$ ($O_1$ term in Table~\ref{tab:sigma_x_model}) and its energy decreases. In contrast, the energy of LUMO$_1$ increases since the ``bonding interaction'' (characterized by $O_1$) is getting smaller. 

In bending along the armchair direction, this decrease/increase of the HOMO/LUMO$_1$ energy is counteracted by contributions from $O_2$. However,  zig-zag bending does not affect $d_2$, so $\Delta \Sigma_X$ shows much higher slopes for both HOMO and LUMO$_1$. 

The overall change of the QP energy ($\Delta \varepsilon^{QP}$) with the curvature ($1/R$) is smaller as shown in Table~\ref{tab:bending}.  This is because of partial cancellation of $\Delta \Sigma_X$ by the changes in $\Sigma_P$, which in all cases studied raises the QP energy. \footnote{The change of $\Sigma_P$ with curvature is $0.85/0.95$~eV~nm and $-0.84/-0.88$~eV~nm for HOMO and LUMO$_1$  along the zig-zag/armchair directions.}

A similar consideration applies also to the LUMO$_2$ states which have distinct character for bending along the zig-zag and armchair axes. In the first case, LUMO$_2$ has an overall anti-bonding $p_x$ character \footnote{As mentioned in Sec.III B, the LUMO$_2$ state appears as anti-bonding only on the outer surface (with respect to the bending axis), while it is bonding on the inner surface. The former interaction dominates since the interatomic distances on the outer surface increase faster (by a factor of $\approx 3.5$) with $1/R$. Hence, the exchange contribution shows overall stabilization with decreasing bending radii; indeed $\Sigma_X$ of $p_x$ state decreases with slope of $-1.37$~eV~nm. }, but we note that $\Delta \Sigma_X / \Delta(\frac{1}{R})$ significantly underestimates the variation of $\varepsilon^{QP}$ ( by  $\sim 50\%$ as shown in Table~\ref{tab:bending}). The remaining part stems from the changes in the Hartree and external potential energies. 

For bending along the armchair direction, LUMO$_2$ has a mixed $p_z$ and $p_x$ bonding character. Due to an increase of $d_1$ and $d_2$ with $1/R$, $\Sigma_X$ increases (i.e., destabilizes LUMO$_2$) with a slope of $0.12$~eV~nm. This is similar to what happens with LUMO$_1$ (but the change is much smaller). This exchange effect is counterbalanced by large changes in $\Sigma_P$ and the electrostatic potential. The LUMO$_2$ QP energy thus slightly decreases with energy. 

Hence, the behavior of the LUMO$_2$ states for bending along the zig-zag and armchair axes has a different origin. While in the first case (zig-zag bending), it is qualitatively given by variation of $\Sigma_X$, the response to bending in the armchair direction is governed by correlations and electrostatic effects. Combined, this leads to a very anisotropic response of the QP energies (and fundamental gaps) to bending.

\begin{table}
\begin{tabular}{l c c c}
\toprule
  &  $R\to\infty$& $R=4$~nm & $R=2$~nm \\
\hline\\
zig-zag  & 2.47 & 2.82 & 2.37\\
armchair& 2.47 & 2.84 & 3.07\\
\bottomrule
\end{tabular}
\caption{Fundamental band-gaps for ideally planar ($R\to \infty$) PNS and two bent systems with radii $R=4$ and $2$~nm along the zig-zag and armchair axes. The stochastic error is 0.04~eV in all cases.}\label{tab:gaps}
\end{table}

\begin{table}
\begin{tabular}{l *{2}{d{3.3}} }
\toprule
  & \mc{$O_1$ [eV nm]} & \mc{$O_2$  [eV nm]} \\
\hline\\
HOMO						& $3.22$ & $-4.63$\\
LUMO$_1$				& $-4.22$ & $5.03$\\
LUMO$_2$ ($p_x$)	& $2.94$ & $-$\\
\bottomrule
\end{tabular}
\caption{Fitted parameters }\label{tab:sigma_x_model}
\end{table}

\begin{table}
\begin{tabular}{l *{4}{d{3.3}} }
\toprule
 & \multicolumn{2}{c}{$\frac{\Delta\Sigma_X}{\Delta(1/R)}$ [eV nm]} &\multicolumn{2}{c}{$\frac{\Delta\varepsilon^{QP}}{\Delta(1/R)}$ [eV nm]} \\
& \mc{zig-zag}&  \mc{armchair}& \mc{zig-zag}& \mc{armchair}\\
\hline\\
HOMO								& -1.67 &-1.24 	& -0.71 &-1.05 \\
LUMO$_1$						& 2.07  &1.20 	&1.05 	&0.44\\
LUMO$_2$ ($p_x$)			& -1.54  &			& -2.91	&\\
LUMO$_2$ ($p_{x+z}$)	&  		&  		&  		&-0.66\\
\bottomrule
\end{tabular}
\caption{Selected slopes (with respect to $1/R$) of the change in the exchange and QP energies for several band-edge states.}\label{tab:bending}
\end{table}

\section{Conclusions}
Ab-initio many-body perturbation theory was used here to study bending-induced changes of $\varepsilon^{QP}$ and band gaps in PNS.  Extremely large PNSs containing $1,958$ valence electrons were studied for bending radii ranging between 1$\mu$m and 2~nm along the armchair and zig-zag directions. Bending along the zig-zag direction shows changes in the QP energies even for very small curvatures (which corresponds to strain $\ll 1\%$) and a bandgap decrease for $R>100$~nm, not observed in the armchair direction. Sample roughness leading  to slight distortion would thus explain variation in experimental $E_g$ as well as apparent in-gap states and peaks in scanning tunneling data~\cite{liang2014electronic}.  

Bending PNS to smaller radii $R<100$~nm results in an opening of the fundamental band gap, regardless of the bending direction. This trend  persists however only till $R\sim4$~nm, at which unoccupied states reorder, leading to a nonmonotonic behavior of the fundamental gap for bending along the zig-zag but not armchair directions. Thus, the behavior of $E_g$ with increasing deformation depends on the direction of the bending and as a result, it is possible to achieve band gap variation as large as $\approx 0.7$~eV within the same material depending only on the bending direction. 

We explained the emergence of the different response to curvature by analyzing individual energy contributions to the quasiparticle levels.  Distinct stability of various unoccupied states was found to derive mostly from exchange terms dominated by the bonding or anti-bonding character of nearest-neighbor orbital overlaps. Variation of QP energies with bending is substantially modified however by dynamical screening, which dominates the response for bending along the armchair direction. For large zig-zag deformations, the first unoccupied state has a $p_x$ anti-bonding character. Its energy quickly decreases with further bending leading to a drop of $E_g$. For the same bending radii along the armchair direction, the first unoccupied state is a hybridized bonding combination of $p_z$ and $p_x$. Due to competing exchange-correlation effects this hybridized state only weakly depends on curvature. Therefore $E_g$ keeps increasing with $1/R$ even for $R<4$~nm for armchair bending.

Results for relaxed bent phosphorene nanosheets corroborate our prediction of LUMO reordering and strong $E_g$ variation depending on the  bending direction.  Hence, bending appears  as a very efficient way to manipulate band gaps and orbital characters in phosphorene. Due to changes in the orbital shape and distribution, such modification could be very useful in understanding and developing optoelectronics and valleytronics devices~\cite{schaibley2016valleytronics}.

\begin{acknowledgement}
D.N. acknowledges support from the NSF Grant No. DMR/BSF1611382. E.R. acknowledges support from the Department of Energy, Photonics at Thermodynamic Limits Energy Frontier Research Center, under grant number DE-SC0019140. R.B. acknowledges support from the US-Israel Binational Science foundation under the BSF-NSF program, Grant No. 2015687. The calculations were performed as part of the XSEDE~\cite{XSEDE} computational Project No. TG-CHE180051.
 The work also used resources of the National Energy Research Scientific Computing Center (NERSC), a U.S. Department of Energy Office of Science User Facility operated under Contract No. DE-AC02-05CH11231. 
\end{acknowledgement}
 
\bibliography{PHPbending}

\end{document}